\documentclass[a4paper]{JHEP3}
\usepackage{epsfig}
\usepackage{graphicx}
\usepackage{amssymb}
\usepackage{latexsym}

\newcommand{\be}{\begin{equation}}
\newcommand{\ee}{\end{equation}}
\newcommand{\bea}{\begin{eqnarray}}
\newcommand{\eea}{\end{eqnarray}}

\newcommand{\al}{\alpha}
\newcommand{\ms}{\rm \overline{MS}}
\newcommand{\f}{\frac}
\renewcommand{\L}{\Lambda}

\title{String-like behaviour of 4d SU(3) Yang-Mills flux tubes}
\author{N.D. Hari Dass
 \\ Institute of Mathematical Sciences, Chennai, India \\ 
 Email: \email{dass@imsc.res.in}}
\author{Pushan Majumdar
\\ Institut f\"ur Theoretische Physik,
Westf\"alische Wilhelms-Universit\"at M\"unster, Germany \\
Email: \email{pushan@uni-muenster.de}}

\abstract{
We present here results on the fine structure of the static 
 $q\bar q$ potential in $d=4$ $SU(3)$ Yang-Mills theory.
The potential is obtained from Polyakov loop correlators having 
separations between 0.3 and 1.2 fermi.
Measurements
were carried out on lattices of spatial extents of about 4 and 5.4 fermi. The temporal
extent was 5.4 fermi in both cases.
The results are analyzed in terms of the force between a $q\bar q$ pair
as well as in terms of a scaled second derivative of the potential.
The data is accurate enough to distinguish between different effective
string models and it seems to favour the expression for ground state
energy of a Nambu-Goto string.
}


\preprint{IMSc/2006/8/20\\ MS-TP-06-9}

\begin{document}

\section{Introduction}

The string picture of hadrons 
\cite{nambu,holger,suss},
has a long history starting from the  
 1960's. 
Although this picture could explain the Regge trajectories of hadrons,
it required 26 space-time dimensions to be a consistent theory \cite{GOD}.
Effective string theories which could be quantized in any dimensions were  
introduced by Polchinski and Strominger in \cite{PS}.
In QCD such strings have the interpretation of gluonic flux tubes between 
a quark and an antiquark. Nambu made formal connections between QCD and 
the string models \cite{N}.

A long distance $1/r$
term in the quark-antiquark potential, distinct from the short distance couloumbic 
$1/r$ term was first observed in \cite{LSW}. Subsequently L\"uscher showed the 
universality of this term and since then it is known as the L\"uscher term \cite{L}, 
having the value $-\pi(d-2)/24r$ where $d$
is the number of space time dimensions and $r$ the length of the string.

In the eighties \cite{ambjorn} and \cite{deF} claimed to find the L\"uscher term in
$d=3$ $SU(2)$ and $d=3+1$ $SU(3)$ lattice gauge theories respectively. 
However precise data on large Wilson loops or Polyakov loop correlators, necessary to
 identify the class of the effective string did not exist at that time.

In recent times, significant progress has been made due to increase in 
computing power as well as improvement in algorithms. 
See \cite{caselle,lw1,pushspec1, pushspec2,caslw,kuticr} for example.

Related to the effective string descriptions is the issue of
the spectrum of string excitations. Recently analytical studies, based only 
on symmetry principles have been carried out in \cite{lweisz,Meyer}. 
Extensive numerical studies have been carried out in 
 \cite{kutispec} and \cite{pushspec1,pushspec2}.
We do not discuss this topic here. One can refer to the review article by
Kuti \cite{kutilat05} for a general introduction and more details.

In this article we present results of our simulations of the Polyakov loop
correlators for $d=4$ $SU(3)$ Yang-Mills theory and compare the resulting 
static potential with both perturbation theory and string model predictions. 
A preliminary report of this simulation was presented in \cite{ourlat05}.

\section{c(r) at short and long distances}
In this article we look at $c(r)=\frac{1}{2}r^3F^\prime(r)$ where
$F(r)$ is the force between a quark and an antiquark. This quantity 
asymptotically tends to the L\"uscher term for large $r$.
We measure $c(r)$ at intermediate distances from slightly 
below 0.5 fermi to about 1.2 fermi on a $32^4$ lattice at $\beta = 5.7$. It 
is therefore interesting to compare the values one obtains  
 from perturbation theory at short distances as well as from the string 
picture. 

$c(r)$ can be easily computed from 2-loop perturbation theory using the results 
in \cite{pcr1} and \cite{pcr2}. We just quote the final result for the sake 
of completeness: 
\be
c(r) = -C_F\left [ \left \{\alpha_{\ms} + c_0 (\alpha_{\ms})^2 \right \} - \frac{r}{2}
\,\partial_r\alpha_{\ms}\left \{ 1+2c_0\alpha_{\ms}\right \}\right ]+\ldots ~.
\ee
Here $C_F=\f{4}{3}$ and $c_0=8\pi \beta_0\left ( \gamma -\f{35}{66}\right )$. 
 The function $\al_{\ms}$ is given by \cite{pcr2}
 \be\label{al}
 \al_{\ms}(r)=\f{1}{4\pi\beta_0f_1}\left ( 1-\f{\beta_1f_2}{\beta_0^2f_1}\right )
 \ee
with $f_1=-\ln\left ( r^2\L^2_{\ms}\right ) $ and 
$f_2=\ln\left (f_1\right ).$ For $SU(3)$ the constants $\beta_0$,
$\beta_1 $ and $\gamma$ have their usual 
continuum values of $\f{11}{(4\pi)^2}$, $\f{102}{(4\pi)^4}$ and 0.57722
 respectively.
The result of this computation is plotted as the curve `2-loop perturbation theory'
in fig \ref{cr:fig} with
the upper and lower curves corresponding to estimated upper and lower limits of
$\L_{\ms}$\cite{pcr1}.

At the other end of the length scale one expects a string like behaviour.
Since non-bosonic forms of the string have been ruled out \cite{LucTep}, 
 our case of interest is the Nambu-Goto string.
The potential in this case was first given by Arvis 
\cite{Ar} to be
 \be\label{Arvispot}
V_{\rm Arvis}(r) = \sqrt{{\sigma}^2r^2-{(d-2)\pi\over 12}\sigma}.
\ee
An interesting point, noted even before Arvis is that the $\bar qq$
 potential becomes purely imaginary for $r < r_c$ where 
$r_c = \sqrt{{(d-2)\pi\over 12\sigma}}$ \cite{Al}, which is around 0.3 fermi.
This behaviour was connected to the tachyon instability of the Nambu-Goto string \cite{Ole}. 
It is curious that perturbative calculations also break down at about the
same distance.

Another form of the potential that is of interest is the so called 
truncated Arvis potential which is obtained by expanding the Arvis potential 
in a power series and retaining the first three terms:
\be\label{trArvis}
V_{\rm trunc} = \sigma r -\frac{(d-2)\pi}{24r}-\frac{(d-2)^2\pi^2}{1152\sigma r^3}.
\ee

\section{Simulation}

\subsection{Algorithm}
Since we are interested in ground state properties, we measure Polyakov 
loop correlators as it is well known that the Polyakov loop correlator 
for large temporal extent strongly projects onto the $q\bar q$ ground state. 
We use the multilevel 
technique of L\"uscher and Weisz for measuring the Polyakov loop correlators
very accurately. For details of the algorithm we refer the reader to \cite{lw1}.

This algorithm has several optimization parameters and the most important 
among them seems to be the number of ``measurements" used to compute certain 
intermediate expectation values. We will refer to this number henceforth 
as ``iupd". Another parameter is the thickness of the time slice. For our 
case a thickness of two was optimal.

It is also well known that improved observables can be constructed by 
replacing a bare link by its group average keeping the environment
unchanged. This is known as multihit and we employ that too on the time-like 
links for the correlators. The group average is computed through the 
single-link integral defined by
\be
\langle U\rangle = Z(J,J^\dag)^{-1}{\partial Z(J,J^\dag)\over\partial J^{\dagger}}\;\;\;\;
{\rm where}\;\;\;\; Z(J,J^\dag) = \int_{SU(3)} [dU]e^{{\rm tr}(UJ^\dag+U^\dag J)}.
\ee
For $SU(3)$ the group average cannot be easily carried out analytically and is
most often evaluated using a monte-carlo method. An alternative semi-analytic 
method for this averaging was proposed by de Forcrand and Roiesnel \cite{forcrand}.
We used the semi-analytic method for the multihit and
we estimate that this resulted in a 60\% speedup of the code compared to
using the monte-carlo method to reach similar levels of accuracy.

In our test runs, for Monte-Carlo multihit the optimal value
for iupd occurred at the same place for the two different values of $r$ we 
considered but not for the 
semi-analytic case. To obtain maximum gain from the multilevel
scheme, we tried to choose iupd close to the optimal values for our
largest $r$.
Another interesting observation is that the penalty paid for
operating at non-optimal values of iupd seems to be higher
for Monte-Carlo multihit than the semi-analytic method.



We have carried out simulations at $\beta=5.7$ on both $24^3\times 32$ and 
$32^4$ lattices using the Wilson gauge action. The lattice spacing at this 
$\beta$ is 0.17 fm so that the temporal extent of the lattice is 5.4 fm
while the spatial box is $(4{\rm \,fm})^3$ in one case and $(5.4{\rm \,fm})^3$ 
in the other. Details about setting the scale is discussed in the next sub-section. 

For separations $r = 2-6$ each measurement involved simulations on 
$24^3\times 32$ lattices with iupd=12000 and 500 measurements were made in all.
For the larger separations $r = 5-9\,$ simulations were done on $32^4$ 
lattices with iupd=48000 and in all again about 500 measurements were obtained.

We also found that for the larger separations 
($r=6$ and higher), we had to go to larger lattices to continue to gain 
from the multilevel scheme. On the $24^3\times 32$ lattice 
increasing ``iupd" even by an order of magnitude did not seem to help reduce 
the error on the $r=6$ correlator. However going to a larger lattice ($32^4$) 
did help significantly. 

\subsection{Results}

\TABLE[t]{\caption{Results: Polyakov loop, Force, $c(r)$\label{tab:res}}
\begin{tabular}{crclcl}
\hline
$r$ & $\langle P^*P\rangle (r)~~~~~$ & $\bar r$ & force & ${\tilde r}$ & $c({\tilde r})$ \\
\hline
2  & $6.683(19)\times 10^{-12}$ & & & & \\
3  & $6.824(38)\times 10^{-15}$ &2.277 & 0.21521(9) & 2.700  & -0.3076(3) \\
4  & $1.894(17)\times 10^{-17}$ &3.312 & 0.18396(11)& 3.729  & -0.3218(8) \\
5  & $7.755(44)\times 10^{-20}$ &4.359 & 0.17155(13)& 4.786  & -0.3141(21)\\
6  & $3.842(30)\times 10^{-22}$ &5.393 & 0.16591(6) & 5.833  & -0.3031(28)\\
7  & $2.098(21)\times 10^{-24}$ &6.414 & 0.16286(10)& 6.864  & -0.302(11)\\
8  & $1.217(17)\times 10^{-26}$ &7.428 & 0.16100(20)& & \\
9  & $7.40(21)\times 10^{-29}$  &8.438 & 0.15975(98)& & \\
\hline
\end{tabular}
}

In our simulations we measure the Polyakov loop correlator $\langle P^*P\rangle(r)$
where $r$ is the separation between the two Polyakov loops. The $q\bar q$ potential
is determined as
\be
V(r)=-\frac{1}{T}\log\langle P^*P\rangle(r)
\ee
where $T$ is the temporal extent of the lattice.
In our analysis we look at the force between the $q\bar q$-pair
given by $F(r) = {d V(r)\over d r}$ and the scaled second derivative $c(r)$
given by $c(r) ={r^3\over 2}{d^2 V(r)\over d r^2}$. 
The latter tends to the
asymptotic value of ${- {(d-2)\pi
\over 24}}$ for large $r$ which is nothing but the L\"uscher term. We are interested in how this quantity
approaches its asymptotic value.
On the lattice these quantities are defined by
\bea\label{force:eq}
F({\bar r})&=&V(r)-V(r-1) \\
c({\tilde r})&=&\frac{{\tilde r}^3}{2}[V(r+1)+V(r-1)-2V(r)]
\eea
where ${\bar r}$ and ${\tilde r}$ are defined as in \cite{lw2} to reduce lattice artifacts.
Our results are presented in table \ref{tab:res}.

Since different $r$ values
are measured in the same simulation, they are quite strongly correlated.
Thus when one evaluates the differences in potentials, it helps if one
evaluates the differences for each measurement and then averages over different
measurements rather than the other way round. In practice we compute the
difference between jackknife bins and then compute the jackknife error
for the differences. We also checked that our combination of one heat-bath with three 
over-relaxation sweeps and restricting the measurement to every fifth update resulted in 
negligible autocorrelation.


\FIGURE[t]{
\includegraphics[width=0.9\textwidth,angle=0]{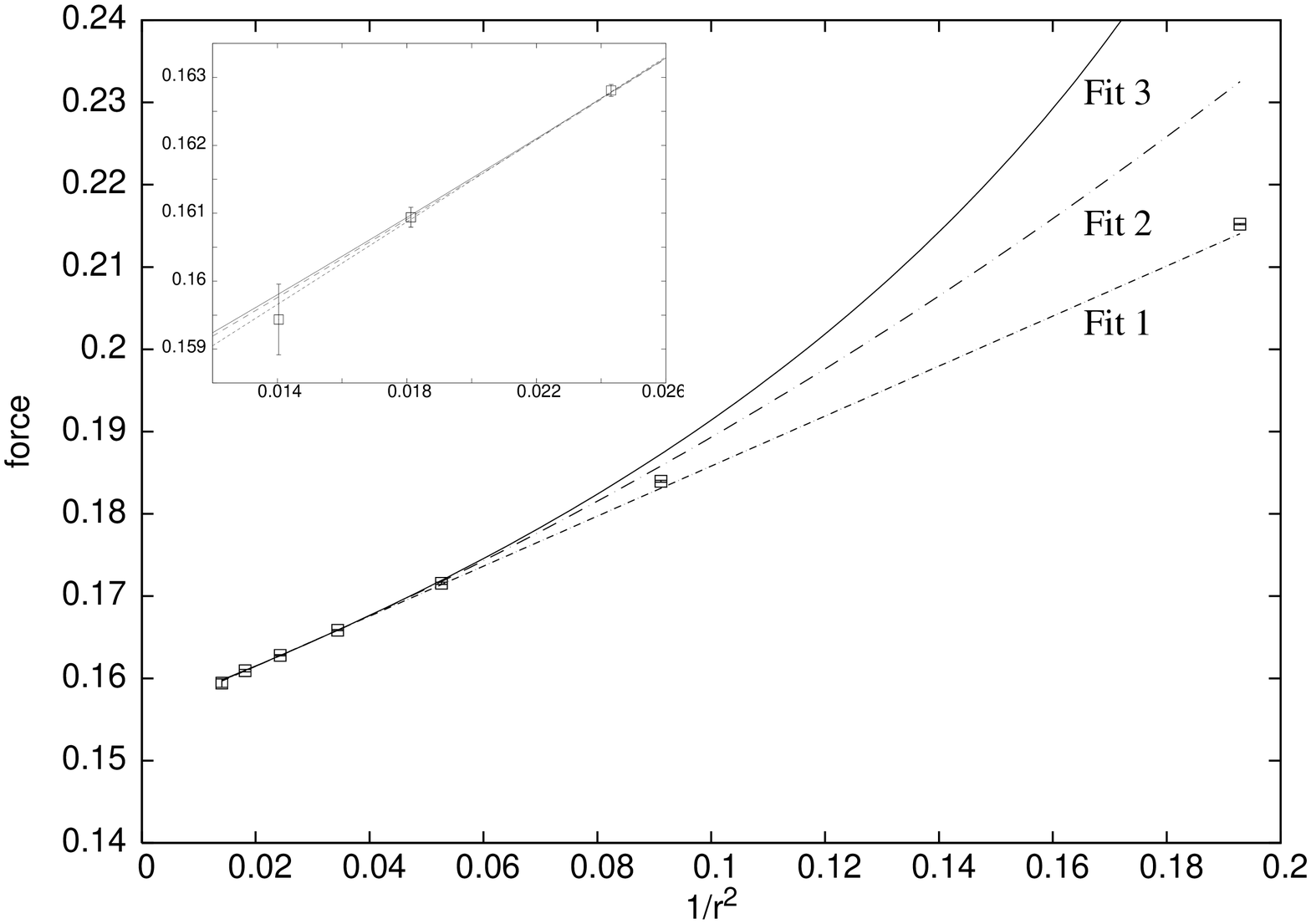}
\caption{Force in $d=4$ $SU(3)$ case. The inset shows the last three points in 
greater detail. The fits 1, 2 and 3 are to a straight line, truncated Arvis and 
full Arvis respectively. Details about the fits are given in table \ref{fit:tab}. }
\label{force:fig}
}

The $q\bar q$ potential contains an unphysical constant which masks the properties 
of the flux tube to a certain degree. We therefore look directly at the force defined 
as in eq. (\ref{force:eq}).  
In figure \ref{force:fig} we have plotted the force versus $1/r^2$. 
We use the force to set the scale on the lattice via $r_0^2F(r_0)=1.65$ where 
$r_0$ is the Sommer parameter which we take to represent the physical distance 
of 0.5 fm. In our case the Sommer parameter turns out to be 2.93 lattice spacings. 

\TABLE{\caption{Fit results\label{fit:tab}}
\begin{tabular}{cccccc}
\hline
\# & functional form & fit range ($r$)& $s$ & $c$ & $\chi^2/d.o.f$ \\
\hline
Fit 1 & $s+c/r^2$ & 6-9 & 0.1554(1) & 0.304(3) & 0.12 \\
& & & & & \\
Fit 2 & $s+c/r^2+3c^2/2sr^4$ & 6-9 & 0.1559(1) & 0.266(3) & 0.27 \\
& & & & & \\
Fit 3 & $s(1-\pi/(6sr^2))^{-\frac{1}{2}}$ & 7-9 & 0.1560(1) & $-$ & 0.3 \\
\hline
\end{tabular}
}

We do three different fits to the force data whose results are 
 shown in table \ref{fit:tab} . 
First we fit  
to the form $sx+c$ where $x=1/r^2$ with $s$ and $c$ as fit parameters. 
From the intercept of this fit we obtain the string tension to be 
 $\sigma a^2=0.1554(1)$ . The slope of this line gives us the effective 
$c(r)$ over this range of $r$ and this turns out to be $0.304(3)$ which is still
16\% away from the asymptotic value of $\pi/12$. This clearly shows that we 
are still some distance away from the region where the model independent leading 
order behaviour is all that matters. 

Next we fit to the form $s+cx+3c^2x^2/2s$ which is inspired by the truncated Arvis      
potential eq. (\ref{trArvis}). This fit gives us the string tension via $s$ as 0.1559(1) 
and the effective 
$c(r)$ to be 0.266(3) which within errors is almost identical to the universal 
value of 
$\pi/12$. Finally we fit to the form expected from the full Arvis potential. 
Within the fit range and errors, this fit is indistinguishable from the fit 
to the truncated Arvis potential.
Unfortunately we cannot yet do an analysis of the the type envisaged in 
\cite{lweisz} where an additional coefficient in front of the $1/r^4$ term can be 
determined by fits.
Such an analysis would probably require 
data with significantly reduced error bars. That seems to be out of scope 
at the moment. What we do see is that the $1/r^3$ term in the potential 
for open strings with fixed end boundary conditions is indeed consistent 
with the Arvis potential and any correction to this would have to be really small.

\FIGURE[!t]{
\centerline{\includegraphics[width=12cm,angle=0]{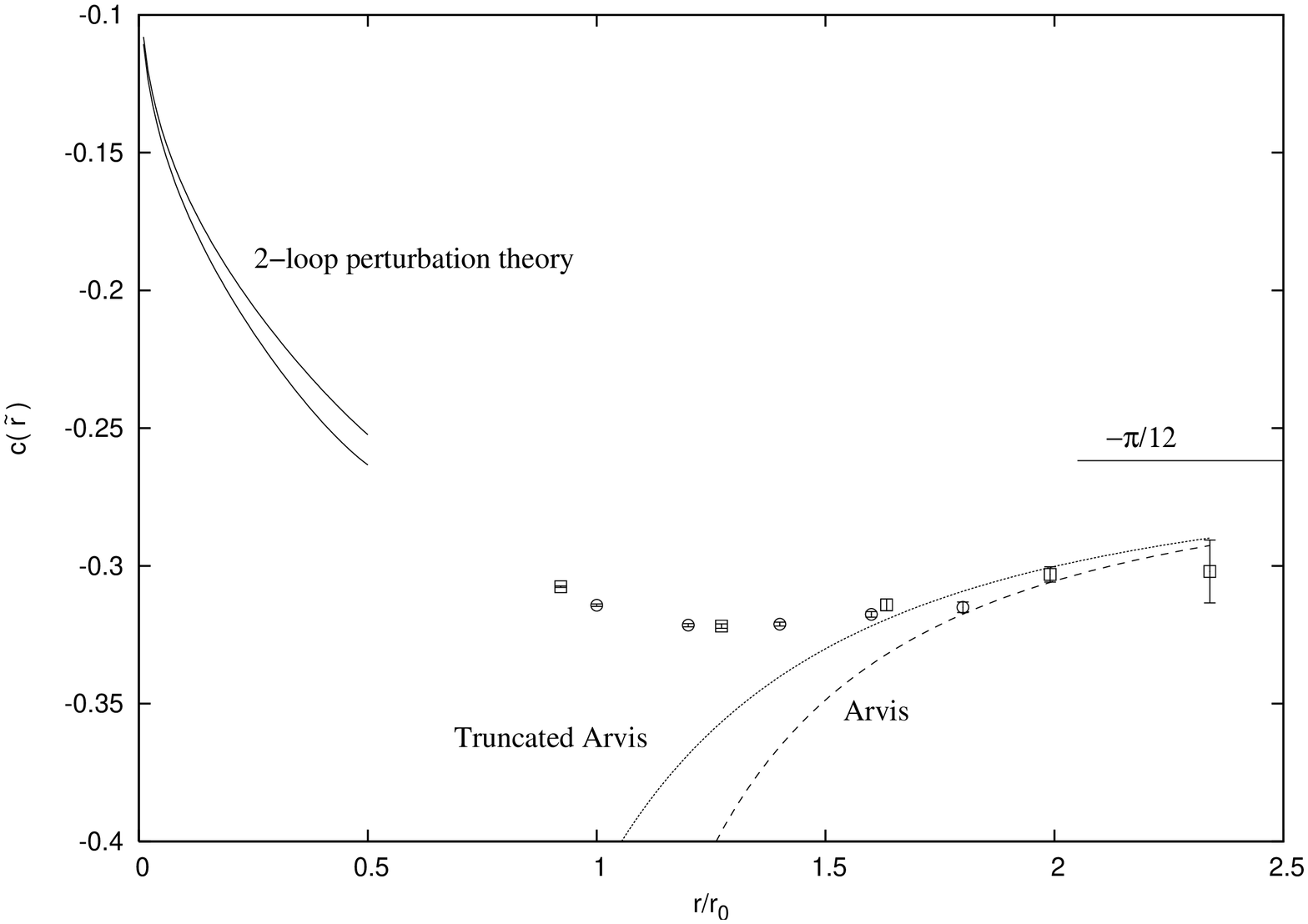}}
\caption{Scaled second derivative for $d=4$ $SU(3)$ case. The circles are from \cite{lw2}}
\label{cr:fig}
}
In figure \ref{cr:fig} we have plotted the scaled double derivative $c({\tilde r})$ 
as a function of $r/r_0$. The horizontal
line at $c(r) = -\pi/12$ is the asymptotic value.
Also shown are the predictions of $c(r)$ from the Arvis as well as the truncated
 Arvis potentials. The perturbative predictions are plotted for small values of 
 $r$ extending upto about 0.25 fermi.

\section{Discussion and conclusions}
In this article we have looked at how the QCD-string behaves at intermediate
distances.The only other simulations of the Polyakov loop correlators in $d=4$ $ SU(3)$ theory
are reported in \cite{lw1,lw2}. To compare our results with them we have plotted their
data along with ours (circles in figure \ref{cr:fig}). In the range where we overlap, the 
agreement of the data is excellent. We have been able to extend the range modestly albeit 
 at a considerable computational cost\footnote{In \cite{lw2} it was estimated that for a 
 given lattice, the computational cost to reach the same level of accuracy in $c(r)$ grows
  as $r^4$ with $r$.}. However even
this extension is important as our two extra points definitely show that the data 
falls quite nicely on the Arvis curve which was not so clear yet with the previous simulations.
We unfortunately cannot really distinguish between the truncated
Arvis potential and the full Arvis potential as the difference between the two curves is
less than 1\% already at $2r_0$ or 1 fermi.

Polchinski and Strominger \cite{PS} found the L\"uscher term as the leading $1/r$ correction
in their effective string theory. 
It has now been claimed (for closed strings) \cite{dru,unp,pwm} that even 
the coefficient of the $1/r^3$ term 
 is the same as the one obtained from the Arvis potential.
 We find it interesting that the same seems to be true for the open strings also.
While in effective string theories there is no apriori reason to expect the full
 Nambu-Goto behaviour, at the moment we do not see any deviations from it beyond a certain
 distance.

This simulation was performed on a comparatively coarse lattice. Existing results on finer 
lattices \cite{lw2} have shown, for the smaller $r$ values, that the data lies above 
the values obtained in this simulation. If that behaviour persists for larger $r$ values too, then the 
scale at which convergence with truncated Arvis potential sets in would be larger. 
It would be really interesting to study the 
continuum limit of this scale. However that is work for the future. 
 
\acknowledgments
One of the authors, PM, gratefully acknowledges the discussions on the perturbative 
calculations with Peter Wiesz and also Philippe de Forcrand for making available
his program for doing the semi-analytic multihit on the links. The simulations 
were carried out on the teraflop Linux cluster KABRU at IMSc as part of the Xth plan project ILGTI.

\end{document}